\documentclass[preprint,showpacs,prl,10pt,twocolumn,showkeys]{revtex4}%
\usepackage{amsfonts}
\usepackage{amsmath}
\usepackage{amssymb}
\usepackage[dvips]{graphicx}%
\begin{document}
\title{Scale dependent superconductor-insulator transition}
\author{D. Kowal and Z. Ovadyahu}
\affiliation{Racah Institute of Physics, The Hebrew University, Jerusalem 91904, Israel }

\begin{abstract}
We study the disorder driven superconductor to insulator transition in
amorphous films of high carrier-concentration indium-oxide. Using thin films
with various sizes and aspect ratios we show that the `critical'
sheet-resistance $R_{{\small \square}}$ depends systematically on sample
geometry; superconductivity disappears when $R_{{\small \square}}$ exceeds
$\approx6~$k$\Omega$ in large samples. On the other hand, wide and
sufficiently short samples of the same batch exhibit superconductivity (judged
by conductivity versus temperature) up to $R_{{\small \square}}$ which is
considerably larger. These results support the inhomogeneous scenario for the
superconductor-insulator transition.

\end{abstract}
\maketitle

\section{Introduction}

Sufficiently strong spatial disorder may precipitate an insulating phase in an
otherwise metallic system. When the ground state of the metal under study is a
superconductor, the disorder-driven transition is called the superconductor to
insulator transition (SIT). The SIT has been studied over many decades,
especially in thin films where tuning the disorder is readily achieved by
varying the film thickness \cite{1,2,3,4,5,6,7,8}. Many common features of the
SIT in such 2D (two-dimensional) systems were found in different
superconducting materials. On the other hand some details of the resistive
transitions such as re-entrant behavior, prompted researchers to focus on
systems that are structurally `homogeneous' presumably with the hope of being
able to characterize the "clean" physics of the SIT. It was however found out
that even purely amorphous materials, that structurally show no irregularities
on scales of $\approx$10 \AA , often exhibit features that are suggestive of
(transport) inhomogeneities. \cite{3}. It was argued that some degree of
inhomogeneity is unavoidable as this is an inherent effect of disorder
\cite{3}. In addition, recent results, most notably a non-monotonous
magneto-resistance has been reported by a number of groups \cite{2,4,5,6,7,8}.
This magnetoresistance involves conductance variation of many orders of
magnitude and is difficult to explain except through a percolation picture
that explicitly treats the system as being inhomogeneous \cite{9}.

In this note we give further evidence for the inhomogeneous nature of
transport near the SIT of amorphous indium-oxide ($In_{x}O$) films. Using
samples with lateral sizes of 1~mm to 0.45~$\mu$m, and aspect ratios in the
range 1 to $\approx$10$^{3}$ we show that the conductivity at liquid helium
temperatures is strongly scale dependent. The results are interpreted as
evidence for long-range potential-fluctuations, and it is suggested that their
presence be taken into account in trying to explain details of the SIT.

\section{Experimental}

\subsection{Sample preparation and measurements techniques}

The present study was done on thin films of amorphous indium oxide $In_{x}O$.
These were prepared in a vacuum chamber capable of achieving a base pressure
of better than 3$\cdot$10$^{-7}$~mbar and equipped with electron-beam and
Knudsen evaporation sources. Oxygen, cleaned by cold trap, could be bled into
the chamber at controlled rates though a needle valve. Standard glass slides
were used as substrates. These were chemically cleaned before being placed in
the chamber. A setup for plasma cleaning of the substrates was used prior to
deposition using pure oxygen. After plasma cleaning for a few minutes in an
atmosphere of 2$\cdot$10$^{-1}$~mbar, the chamber was pumped down to better
than 5$\cdot$10$^{-6}$ mbar. Films were prepared by e-beam evaporation of
99.999\% pure pressed In$_{2}$O$_{3}$ pieces manufactured by Cerac as
described in detail elsewhere \cite{10}. A quartz crystal thickness monitor
(calibrated against a Tolanski interferometer) was used to determine
deposition rates and film thickness. The $In_{x}O$ films used in this study
were deposited at a rate of 1 \AA /s to a nominal thickness of 200-250~\AA .
These conditions yield the relatively high carrier concentration
$n\gtrsim5\cdot10^{21}~cm^{-3}$ ($In_{x}O$ samples with $n$ smaller than
$\simeq10^{20}~cm^{-3}$ are not superconducting down to 1~K). The as-prepared
films were studied by high resolution transmission electron microscopy and
exhibited microstructure and diffraction patterns that are typical for
amorphous structures (c.f., figure~1).

Samples for measuring large scale material properties and Hall-effect
measurements were defined by placing an appropriate aluminum mask over the
substrate during the deposition. To study the scale dependence, we compared
the properties of batches of simultaneously prepared samples of common width
and different lengths down to less than a micron. These were prepared by a
liftoff technique; Strips of amorphous indium-oxide, 0.5$~$mm wide and few
millimeters long were made using a mechanical mask. Thin fibers were then
pulled from a drop of 7031-GE varnish and placed across the strip. The
thickness of the fibers could be controlled in the range of $\simeq$50$~\mu$m
down to less than 0.2$~\mu$m by proper dilution of the varnish in a 1:1
solution of ethanol and toluene (varnish thinner), and by the speed at which
the fibers were pulled. The slides were then returned to the vacuum chamber
and a $\approx$300~\AA ~layer of gold was deposited over the $In_{x}O$.
Finally the slides were immersed in varnish thinner to dissolve the fibers and
wash away the gold that was on them. This left a series of gaps in the gold
layer which defined indium oxide samples all "cut" from the same thin film,
having a common width but different lengths. Longer samples could be made by
using commercial copper wire instead of the varnish fibers. Electrical
connections were made by soldering wires to small pieces of indium metal
pressed onto the gold contacts. Before depositing the gold and after
completing sample preparation the fibers/gaps were viewed through an optical
microscope at x400 magnification to see the range of sample lengths obtained
and their geometrical quality. Figure 2 shows a SEM micrograph of a 0.9$~\mu$m
long sample. The borders were checked to bee smooth and equidistant over the
entire width of the sample (500$~\mu$m) to within $\approx$10\%.%
\begin{figure}
[ptb]
\begin{center}
\includegraphics[
height=2.1181in,
width=2.9386in
]%
{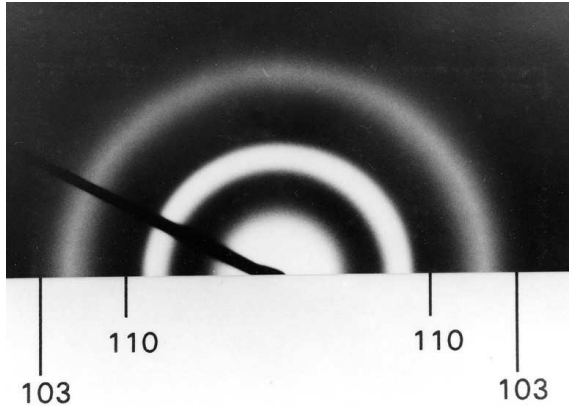}%
\caption{Electron diffraction pattern of as-made $In_{x}O$ film showing only
amorphous rings, in particular, the $<$110$>$ and $<$103$>$ diffraction rings
of the crystalline version are not observed.}%
\end{center}
\end{figure}

\section{Results and discussion}

As prepared films of $In_{x}O$ are usually insulating and exhibit negative
temperature coefficient of resistivity from room temperatures down, and some
form of variable range hopping may be observed at the 10-100~K range
\cite{10}. Samples that are near the SIT on the insulating side tend to show
simple activation at liquid helium temperatures. Namely, $R_{_{{\small \square
}}}(T)$ is seen to follow $R_{_{{\small \square}}}(T)\propto\exp[-\frac{T_{0}%
}{T}]$ with typical activation energy $T_{0}$ of order 3-15~K depending on
disorder. This is true in 3D \cite{11} as well as in the 2D regime
\cite{3,4,5,6,7,8} and is still one of the important unsolved issues of
transport near the SIT \cite{12}. Figure 3 illustrates this behavior for a
large-area film measured at several stages of annealing (disorder). Note that
as the sheet-resistance $R_{_{{\small \square}}}$ at $T$ $\cong$ 4~K is
reduced by annealing, so does the activation energy, and for
$R_{_{{\small \square}}}\approx$6~k$\Omega$ the temperature dependence becomes
rather weak. In fact, as shown in the inset, samples with this value of
$R_{_{{\small \square}}}$ are superconducting below $\approx$0.3~K. A
`critical' $R_{_{{\small \square}}}$ may be thus identified at the value of
$R_{_{{\small \square}}}^{c}\approx$6~k$\Omega$ for the SIT in this system. A
similar value for $R_{_{{\small \square}}}^{c}$ has been reported by other
researchers.%
\begin{figure}
[ptb]
\begin{center}
\includegraphics[
height=2.4569in,
width=2.4729in
]%
{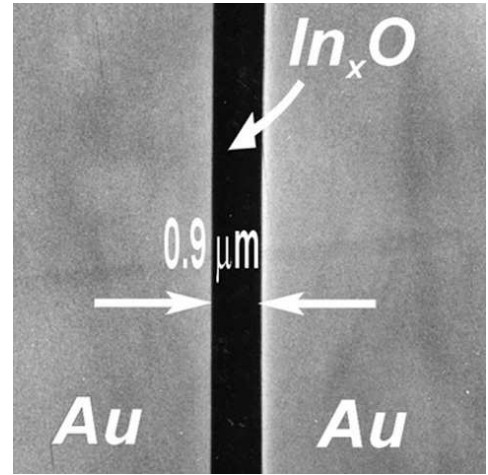}%
\caption{SEM micrograph of a 0.9 $\mu$m long amorphous sample. The width is
500 $\mu$m (extends far beyond the range shown in the micrograph).}%
\end{center}
\end{figure}

A somewhat different situation is encountered in the small-area samples. This
is illustrated in figures 4a and 4b that show $R_{_{{\small \square}}}(T)$
data for a single batch of samples at two stages of annealing and for samples
with different length $L.$ Note that $T_{0}$ has a marked dependence on $L$
(and/or the aspect ratio) and this leads to a significant variation of
$R_{_{{\small \square}}}$ as the temperature is lowered (figure 4a). Upon
further annealing (causing a decrease in the average disorder), this disparity
makes for a sharper division of the $R_{_{{\small \square}}}(T)$ data; The two
longest samples in the batch are still activated with $T_{0}$ somewhat reduced
relative to the previous stage. However, the two shortest samples of the same
batch now show a clear tendency to become superconducting at low temperatures
(figure 4b). This is a striking result as the samples are all made from the
same batch and in fact show very similar resistances at higher temperature.
Moreover, associating, as previously done, $R_{_{{\small \square}}}^{c}$ with
the value of the sheet resistance that separates these two groups we now get a
value that is at least twice larger than that of the large-area samples of the
same material. It is emphasized that the length-controlling fibers were placed
on the common $In_{x}O$ strip in close proximity to one another and at no
particular order (\textit{i.e}., no hierarchy of $L$ along the strip), and
therefore the $L$ dependence cannot be related to a `technological' inhomogeneity.

These dependencies of the transport properties on sample scale and aspect
ratio is suggestive of a percolation phenomena in an inhomogeneous media. As
argued before \cite{3} the inhomogeneity is a natural consequence of disorder,
and inasmuch as superconductivity is concerned, the effect of disorder may be
highly accentuated.%

\begin{figure}
[ptb]
\begin{center}
\includegraphics[
trim=0.120220in 4.532159in 0.247751in 0.407999in,
height=2.8366in,
width=3.2827in
]%
{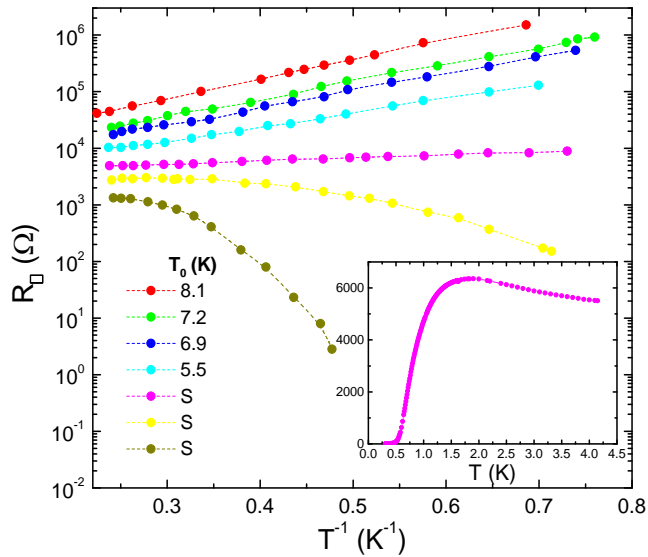}%
\caption{Resistance versus temperature for a large area (500x500 $\mu$m$^{2})$
$In_{x}O$ film at various stages of the annealing process (labeled with their
activation energies or by "S" when superconducting in range). Inset shows a
similar film measured to lower temperatures revealing its superconductivity.}%
\end{center}
\end{figure}

The argument is based essentially on the phenomenon of universal conductance
fluctuation: Superconductivity usually vanishes when the average conductance
of the 2D system is comparable with $e^{2}/h.$ At this degree of disorder the
conductance of the underlying diffusive metal fluctuates considerably on
scales of order $L_{\phi},$ the phase-coherent length. In fact, the local
conductivities $g(L_{\phi})$ are naturally distributed over a $\pm e^{2}/h$
range, which means that at the transition to strong localization the
distribution-width of the local conductances is of order unity. Therefore as
the average critical disorder is attained, there are perforce some regions of
the sample where the local disorder is smaller than critical and these regions
may be locally superconducting. It is easy to see that from the point of view
of conductivity the system may appear `granular' with small superconducting
regions embedded in an insulating matrix (and a complementary situation on the
`weakly localized' side of the transition). This may happen even in a
structurally featureless amorphous structure as are the samples in the present
case (as mentioned above, c.f., figure 1). Such a scenario is generic and
should be pertinent for all materials where sub-critical disorder does not
appreciably decrease $T_{C}$. These then include all systems that obey the
Anderson Theorem as well as metals that show `enhancement' phenomena.%
\begin{figure}
[ptb]
\begin{center}
\includegraphics[
trim=0.238004in 0.831109in 0.670147in 1.586662in,
height=3.6774in,
width=2.8898in
]%
{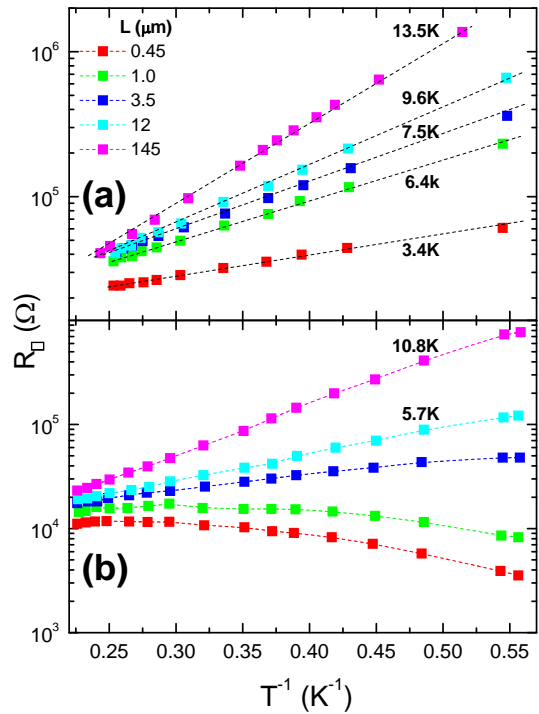}%
\caption{\textbf{(a) }Resistance versus temperature for a single batch of
$In_{x}O$ film with a common width of 500 $\mu$m and length as shown. Each
sub-sample is labeled with its activation energy $T_{0}$ (see
text).\textbf{(b) }Same batch of samples as in (a) after further thermal
annealing.}%
\end{center}
\end{figure}
%

\begin{figure}
[ptb]
\begin{center}
\includegraphics[
trim=0.163272in 4.396159in 0.202263in 0.339418in,
height=2.8508in,
width=3.2038in
]%
{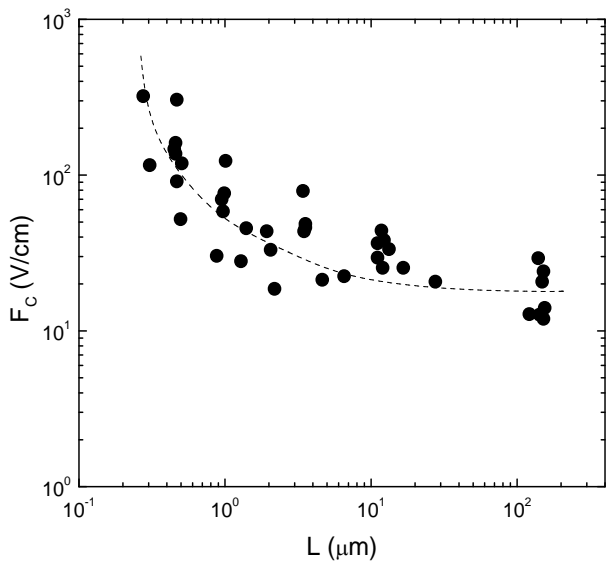}%
\caption{The electric field $F_{c}$ that reduces the resistance by 1\%
(relative to the linear response value) as function of the sample length~$L.$
The dashed line is merely a guide to the eye.}%
\end{center}
\end{figure}

If one allows for just short-range disorder, these arguments cannot however
account for the spatial scale over which the results in figures 4a and 4b are
still size dependent. Gaussian disorder could conceivably lead to an
inhomogeneous situation on scale of the order of $\approx L_{\phi}$ evaluated
at the measurements temperatures (or the percolation radius $\mathcal{L}_{c}$
of the hopping problem on the insulating side of the transition). Neither of
these lengths is larger than 1~$\mu$m at liquid helium temperatures whereas
significant size dependence is observed in our experiments up to $L$ that is
at least one order of magnitude larger.

The relevant scale for the actual `granularity' may of course be larger than
the size of the individual superconducting island; several such islands may be
Josephson-coupled to form a larger cluster. But this still leaves unanswered
the scale dependence of the activation energy observed in figure 4a to extend
over tens of microns. It would therefore seem necessary to consider the
existence of long-range potential-fluctuations in these samples, which is not
an uncommon phenomenon in amorphous materials \cite{13}. In fact, long-range
fluctuations in the disorder may be expected to be present when the disorder
is large as must be the case here (strong disorder is actually required when a
high carrier-concentration system is to be rendered Anderson localized). This
is due to the cost of creating a potential gradient which would favor large
potentials drops on the longer spatial scales. Naturally then, long range
potential fluctuations should be a consideration especially when the sample is
prepared or is later exposed to room (or higher) temperatures.

The presence of long-range potential-fluctuations in these films is also
reflected in the current-voltage ($I-V$) characteristics of the studied
samples. In their normal state all our samples showed deviations from Ohmic
behavior above a characteristic voltage $V_{C}$. These deviations were always
such that $\partial R/\partial V<0,$which is presumably related to the fact
that the temperature coefficient of the resistance is negative at helium
temperatures. As a rule, the critical field $F_{C}=V_{C}/L~$at which
non-linearity became important was noticeably larger at small $L.~$This trend
is illustrated in figure 5 for all the relevant samples used in our study.
Despite the scatter in the data, it is clear that long samples are "softer"
than short ones as may be expected from long-range fluctuations; samples with
larger $L$ are associated with with larger local fields hence they are more
susceptible to non-linearity at any macroscopic field. Note that $F_{C}$~at
which measurable deviations from Ohm's law are observed is scale-dependent
over the same range of $L~$where the data in figure 4a and 4b exhibit scale-dependence.

In summary, we have presented experimental evidence for the inhomogeneous
nature of transport in an amorphous system near the SIT. The results seem to
suggest the relevance of long-range potential-fluctuations which may result
from thickness and/or composition variations to which characterization methods
such as diffraction, STM, TEM, as well as other form of microscopy may not be
as sensitive as charge transport. These issues clearly deserve further
experimental and theoretical elucidation.

This research was supported by a grant administered by the US Israel
Binational Science Foundation and by the Israeli Foundation for Sciences and Humanities.

\end{document}